\documentclass[%
 aip,
 jmp,%
 amsmath,amssymb,
 reprint,%
]{revtex4-1}

\usepackage{graphicx}
\usepackage{dcolumn}
\usepackage{bm}
\usepackage{epstopdf}
\usepackage{amssymb}
\usepackage{color}
\usepackage[colorlinks=true, letterpaper=true, pdfstartview=FitV, linkcolor=blue, citecolor=blue, urlcolor=blue]{hyperref}
\usepackage{lineno}

\begin{document}

\preprint{AIP/123-QED}

\title[]{Strong phonon anharmonicity and low thermal conductivity of monolayer tin oxides driven by lone-pair electrons}

\author{Wenhui Wan}
\author{Yanfeng Ge}
\author{Yong Liu}
\email{ycliu@ysu.edu.cn or yongliu@ysu.edu.cn}
\affiliation{State Key Laboratory of Metastable Materials Science and Technology $\&$
Key Laboratory for Microstructural Material Physics of Hebei Province,
School of Science, Yanshan University, Qinhuangdao, 066004, P.R. China}

\begin{abstract}
  Motivated by the excellent electronic and optoelectronic properties of two-dimensional (2D) tin oxides, we systematically investigated the thermal conductivity ($\kappa$) of monolayer SnO and SnO$_{2}$ by the first-principles calculations. The room-temperature $\kappa$ of monolayer SnO and SnO$_{2}$ reaches 9.6 W/(m$\cdot$K) and 98.8 W/(m$\cdot$K), respectively. The size effect is much weaker for monolayer SnO than for monolayer SnO$_{2}$, due to the coexistence of size dependent and independent component in the $\kappa$ of monolayer SnO. The large difference between the $\kappa$ of 2D tin oxides can be attributed to the small phonon group velocity and strong anharmonicity strength of monolayer SnO. Further electronic structure analysis reveals that the existence of sterically active lone-pair electrons is the key factor for the small $\kappa$ of monolayer SnO. These results provide a guide for the manipulation of thermal transport in the electronic or thermoelectric devices based on 2D tin oxides.
\end{abstract}

\pacs{63.20.-e 63.22.-m, 68.60.Dv, 63.20.kg}
\keywords{Two-dimensional, Tin oxides, Thermal conductivity, lone-pair electrons}
\maketitle


Two-dimensional (2D) materials have attracted much attention as the candidates for developing next-generation high-performance electronics, optoelectronics, and spintronic devices.~\cite{Novoselov2005,chhowalla2013,Nicolosi1226419} However, prototype 2D materials such as graphene, transition metal dichalcogenides (TMDs) and phosphorene are usually unstable at high temperature or oxidize gradually in the air.~\cite{Phosphorene1,MoS2O} 2D metal oxides, by contrast, are environmentally stable, relatively easy to fabricate and typically comprised of non-toxic, naturally occurring
elements.~\cite{SnOproduce1} Thus, many efforts have been denoting into the synthesis, fundamental properties and devices applications of metal oxides.~\cite{synthesis,Tokura462,SnOthermal}

Tin oxides are of considerable technological interest as a series of metal oxides. Under atmospheric conditions, tin oxides usually consist of both SnO and SnO$_{2}$.~\cite{DEPADOVA1994379,SnOSnO2}
Tin oxides have shown high promise in the field of gas sensing,~\cite{HIEN2014134,KUMAR2009587} field effect transistors,~\cite{SnOFET,SnO2FET} as anode materials,~\cite{SnOanode,SnO2anode} thermoelectrics~\cite{SnOthermal} and optoelectronic devices.~\cite{Snoele}
Besides, SnO can exhibit bipolar conductivity under suitable conditions which is of great potential in the design of p-n junction.~\cite{Nomura2011} Meanwhile, transition metal doped SnO$_{2}$ was identified as dilute magnetic semiconductors with curie temperature $T_{c}$ close to 650 K.~\cite{SnO2mage}
2D materials usually possess unique properties that differ from their bulk counterparts. The optical and electrical properties of SnO nanosheets were demonstrated to be strongly dependent on its dimensions.~\cite{SnOoptical} Bulk SnO is a layered material with a tetragonal PbO-type crystallographic structure. Recently, monolayer SnO has been synthesized.~\cite{SnOproduce,SnOproduce1} Ferromagnetism was predicted in the hole-doped monolayer SnO, due to its valence band which has a Mexican-hat band edge.~\cite{magneti} On the other side, though bulk SnO$_{2}$ has a rutile crystal structure, porous monolayer SnO$_{2}$ nanofilm has been fabricated,~\cite{SnO2produce} indicating that the growth of 2D SnO$_{2}$ is likely to be realized in near future. A T-phase structure was predicted as the ground-state structure of monolayer SnO$_{2}$.~\cite{SnO2predict} The magnetic properties of Co-doped and transition metal doped monolayer SnO$_{2}$ have been investigated.~\cite{SnO2mage2,SnO2mage3}
Compared to the intense research on the synthesis and electronic properties of 2D tin oxides, their thermal properties receive less attention.
Thermal conductivity ($\kappa$) is a fundamental physical quantity which has a large influence on many applications such as the heat dissipation of the integrated electronic devices and the efficiency of thermoelectric conversion.~\cite{zou2001,silinano1}
The thermal transport properties of 2D tin oxides have not been measured in experiment up to date and call for a systematical investigation at a microscopic level.

In this work, we calculated the thermal conductivity of monolayer SnO and SnO$_{2}$ by the first-principles calculations coupled with the phonon Boltzmann transport equation (BTE).
The stable structures of 2D tin oxides are identified. Monolayer SnO$_{2}$ have a high room-temperature $\kappa$ which is dominated by acoustic phonons. In contrast, monolayer SnO has a $\kappa$ which is almost ten times lower than that of SnO$_{2}$. The size effect of $\kappa$ in tin oxides was investigated. We analyzed the role of phonon group velocity, anharmonicity and phase space for anharmonicity scattering in the thermal transport. The imparity of $\kappa$ between 2D tin oxides can be attributed to the lone-pair electrons in the monolayer SnO.


The lattice thermal conductivity $\kappa$ is obtained by~\cite{li2014shengbte}
\begin{eqnarray} \label{kappa}
\kappa_{\alpha\beta}= \frac{1}{N\Omega}\sum\limits_{
\mathbf{q},s} {C_{\mathbf{q},s}v_{\mathbf{q},s}^{\alpha}v_{\mathbf{q},s}^{\beta}\tau_{\mathbf{q},s}},
\end{eqnarray}
where $N$ and $\Omega$ are the number of $\mathbf{q}$ point and volume of the unit cell, respectively. $C_{\mathbf{q},s}$, $v_{\mathbf{q},s}^{\alpha}$ and $\tau_{\mathbf{q},s}$ is the mode specific capacity, group velocity along the $\alpha$-th direction and phonon lifetime in the single-mode relaxation time approximation (RTA) of the phonon with wavevector $\mathbf{q}$ and branch index $s$, respectively. $\tau_{\mathbf{q},s}$ was obtained by combing the anharmonic scattering, isotopic impurities scattering and boundary scattering, according to the Matthiessen rule~\cite{li2014shengbte}
\begin{eqnarray} \label{tau}
\frac{1}{\tau_{\mathbf{q},s}}=\frac{1}{\tau_{\mathbf{q},s}^{an}}+\frac{1}{\tau_{\mathbf{q},s}^{iso}}+\frac{1}{\tau_{\mathbf{q},s}^{b}},
\end{eqnarray}
where the completely rough boundary with size of $L$ was used, leading to $1/\tau_{\mathbf{q},s}^{b}=L/|v_{\mathbf{q},s}|$.
All computational details are given in the supplementary material

\begin{figure}[tbp!]
\centerline{\includegraphics[width=0.45\textwidth]{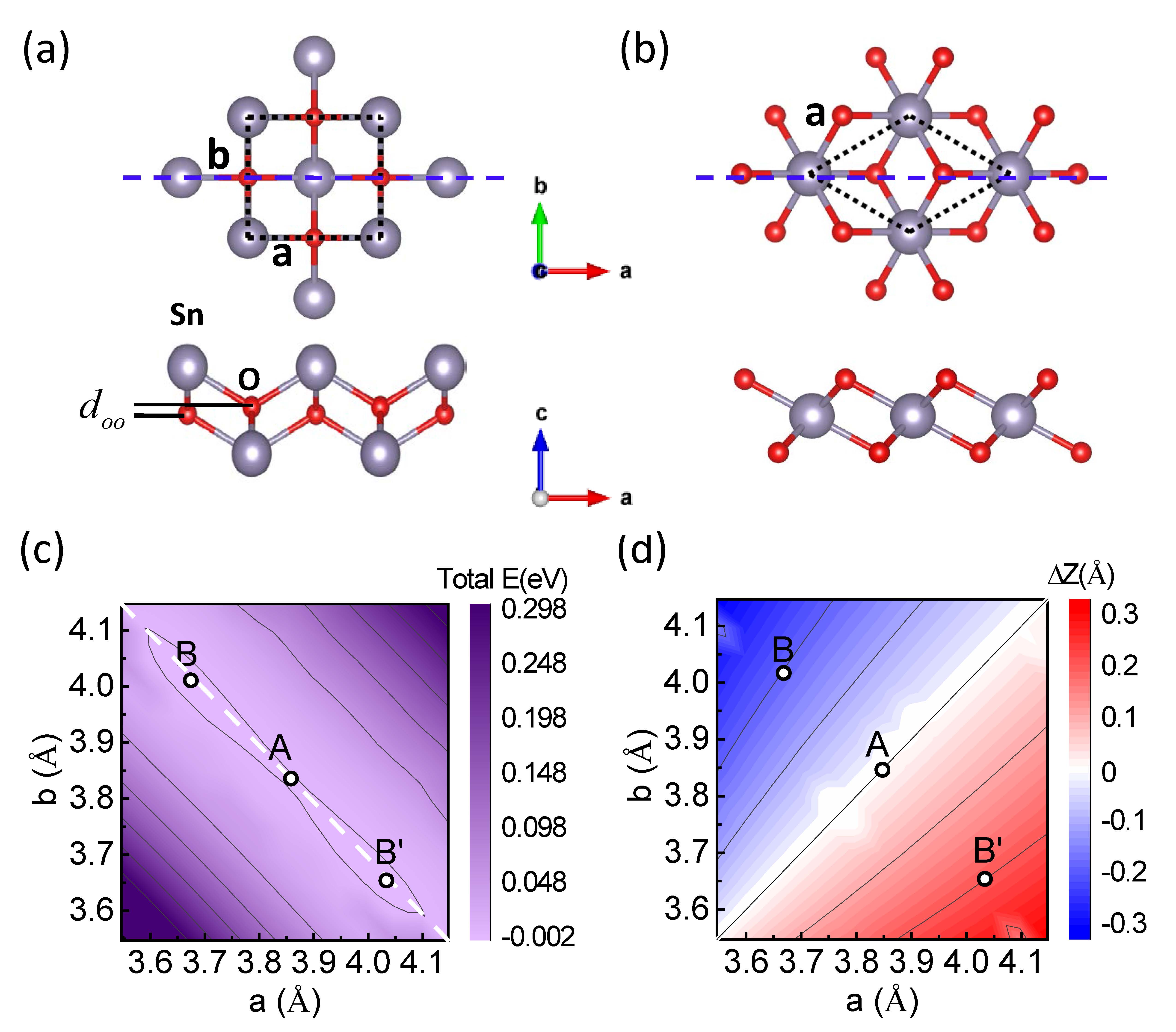}}
\caption{A side view and a top view of crystal structure of (a) monolayer SnO and (b) SnO$_{2}$ with the bash line showing the unit cell. (c) The free-energy contour of SnO as function of lattice constant $a$ and $b$. (d) The distance between O atoms $d_{oo}$ along the out-of-plane direction as a function of $a$ and $b$. }
\label{wh1}
\end{figure}

Figure~\ref{wh1}(a) displays the lattice crystal of monolayer SnO, with two Sn and O atoms in a unit cell. $a$ and $b$ is the lattice constant along $x$- and $y$-axis, respectively.
Unlike bulk and bilayer SnO, monolayer SnO with $a=b=3.847$ \AA\ (labeled as phase A) is unstable (see Fig. S2(a-c) in the supplementary material), agreeing with previous works.~\cite{C5CP02255J,Mounet2018}
The eigenvectors of phonon modes with imaginary frequencies correspond to the anti-directional movement of O atoms along the $z$-axis.
The contour of total energy of monolayer SnO, as the function of $a$ and $b$, is presented in Fig.~\ref{wh1}(c). Compared to phase A, another rectangle structure is more stable which has $a=4.022$ \AA\ and $b=3.663$ \AA, or vice versa (labeled as phase B or B$^{'}$). For more clarity, Fig. S2(d) in the supplementary material displays the energy profile along the path B-A-B$^{'}$ (see bash line in Fig.~\ref{wh1}(c)). Meanwhile, the distance between O atoms ($d_{oo}$) along the $z$ direction changes from zero in phase A to 0.2 \AA\ in phase B (see Fig.~\ref{wh1}(d)). That is consistent with aforementioned eigen-displacement of the imaginary modes in phase A.
Moreover, the structural instability of phase A was further identified with different simulation methods (see Fig. S2(d) in the supplementary material). For the case of monolayer SnO$_{2}$, we adopted T-phase structure~\cite{SnO2predict} (see Fig.~\ref{wh1}(b)). The thickness of monolayer SnO$_{2}$ is chosen by the interlayer distance of bilayer SnO$_{2}$ with the most stable structure (see Fig. S3 in the supplementary material). All the optimized structural parameters are listed in Table.~\ref{table1}.
The crystal symmetry of crystal lattice of monolayer SnO is $D_{2h}$ while that of monolayer SnO$_{2}$ is $D_{3d}$ which contains an inversion center on Sn site.
The bonding length of Sn-O is longer in monolayer SnO than that in monolayer SnO$_{2}$.

\begin{table}[b!]
\caption{\label{table1}
The lattice constant ($a,b$), Sn-O bonding length ($d_{Sn-O}$) and thickness ($l$) of monolayer tin oxides.}
\begin{ruledtabular}
\begin{tabular}{cccccccc}
              &$a,b$(\AA)    &$d_{SnO}$(\AA)  &$l$ (\AA) \\
\hline
SnO           &4.022,3.663 &2.281, 2.228    &4.826     \\
SnO$_{2}$     &3.227       &2.121 	        &4.272     \\
\end{tabular}
\end{ruledtabular}
\end{table}

\begin{figure}[tbp!]
\centerline{\includegraphics[width=0.45\textwidth]{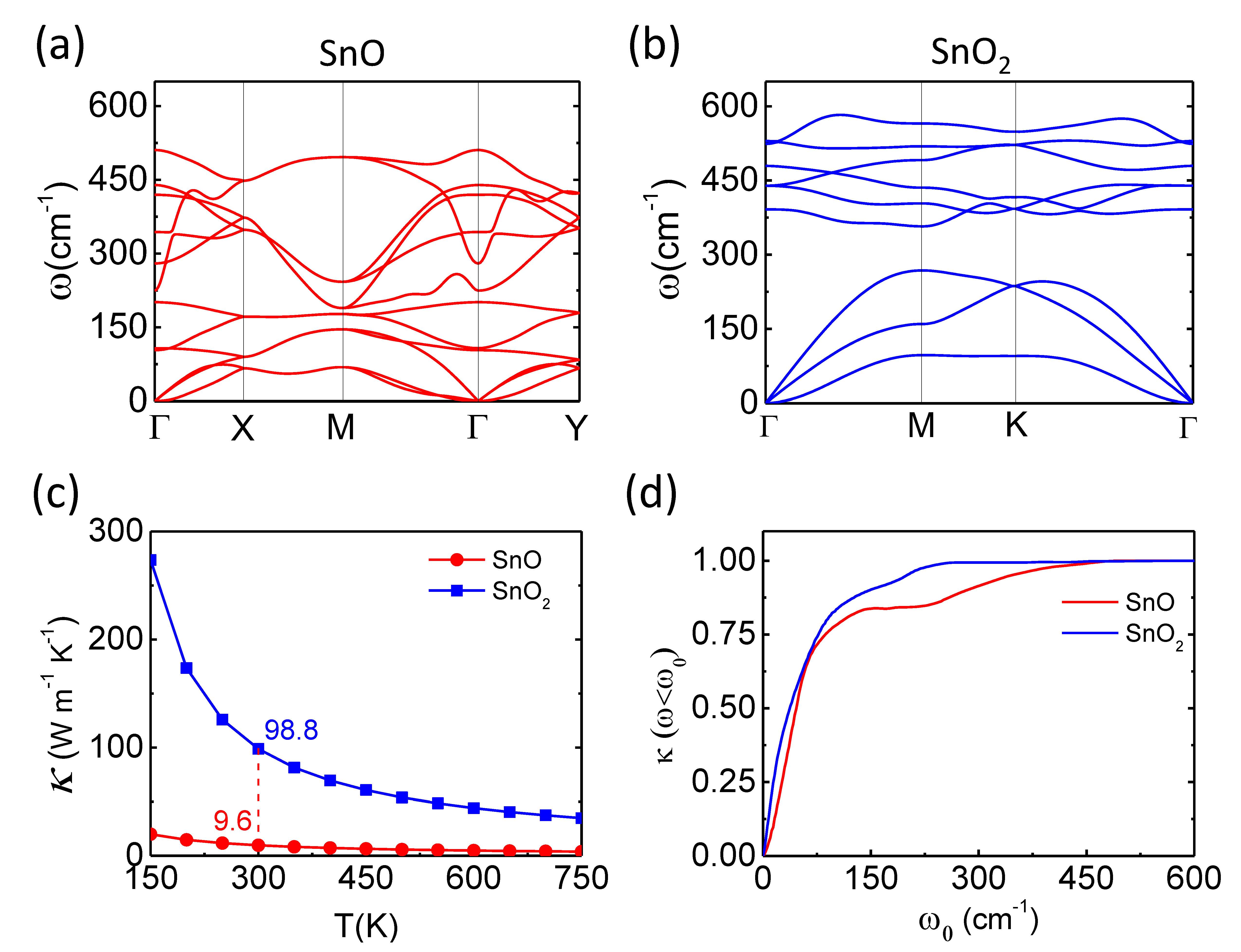}}
\caption{The phonon dispersion of (a) monolayer SnO and (b) monolayer SnO$_{2}$. (c) Temperature dependence of $\kappa$ of monolayer SnO and monolayer SnO$_{2}$. (d) At room temperature, the frequency dependence of the normalized cumulative $\kappa(\omega < \omega _{0})$ for monolayer SnO (red) and SnO$_{2}$ (blue).}
\label{wh2}
\end{figure}

Pristine 2D tin oxides are non-magnetic semiconductors (see Fig. S4 in the supplementary material) and heat is mainly carried by phonons. The phonon dispersions of monolayer SnO in phase B and monolayer SnO$_{2}$ are displayed in Fig.~\ref{wh2}(a,b), which ensure their structural stability. The lowest three branches are the out-of-plane acoustic (ZA), transverse acoustic (TA) and longitudinal acoustic (LA) branch.
ZA branch has a quadratic dispersion near the $\Gamma$ point.~\cite{carrete2016physically}
An acoustic-optical gap is absent and present in the phonon dispersion of monolayer SnO and SnO$_{2}$, respectively.
The low- and high-frequency phonon modes are mainly dominated by the vibration of Sn and O atoms, respectively (see Fig. S5 in the supplementary material).

The temperatures (T) dependence of $\kappa$ without boundary scattering is shown in Fig.~\ref{wh2}(c). At room temperature, the $\kappa$ of monolayer SnO$_{2}$ reaches 98.8 W/(m$\cdot$K), comparable with that of conventional semiconducting Ge (65 W/(m$\cdot$K))~\cite{Gekappa} and GaAs (45 W/(m$\cdot$K)).~\cite{Gekappa} Though the lattice crystal of monolayer SnO is anisotropic, its $\kappa$ is almost isotopic.
The small anisotropy of $\kappa$ is keeping pace with that of phonon group velocity,~\cite{C6NR01349J} which can be further attributed to the similar atomic bonding characteristics along the $x$- and $y$-direction (see Fig. S6 in the supplementary material).
It is possible to induce the anisotropic thermal transport into 2D tin oxides by dislocation~\cite{sun2018dislocation} or uniaxial strain~\cite{uniaxial1} for broadening their potential applications in future.
We will only discuss the $\kappa$ along the $x$-axis hereafter. The room-temperature $\kappa$ of monolayer SnO, however, is only 9.6 W/(m$\cdot$K) which is ten times smaller than that of SnO$_{2}$. Such a large imparity of $\kappa$ is consistent with the large difference between the $\kappa$ of bulk SnO and SnO$_{2}$ observed in experiment.~\cite{Shimpei}
Therefore, thermal transport is the limit factor in electronic applications of 2D SnO rather than 2D SnO$_{2}$.

By fitting the phonon density of states (DOS) at low phonon frequencies with a Debye model,~\cite{phonopy} we obtained Debye temperature $\theta_{D}$ as 259 K and 356 K for monolayer SnO and SnO$_{2}$, respectively. As a result, the $T$ dependence of $\kappa$ of monolayer SnO$_{2}$ follows the relation of $\kappa \sim e^{-\theta_{D}/\beta T}$ at temperature lower than $\theta_{D}$.~\cite{ziman} Here $\beta$ is an empirical parameter which was fitted as 1.02 (see Fig. S7(d) in the supplementary material). At temperature substantially beyond $\theta_{D}$, the overall $T$ dependence of $\kappa$ of 2D tin oxides restores the $1/T$ behavior (see Fig. S7 in the supplementary material).

We found that the effect of isotopic impurities scattering to $\kappa$ is small compared with the anharmonic scattering in the temperature above 250 K. For example, if we exclude the $1/\tau_{\mathbf{q},s}^{iso}$ from the total $1/\tau_{\mathbf{q},s}$ in Eq.~\ref{tau}, the room-temperature $\kappa$ will increase only by 0.6\% and 3.3\% for monolayer SnO and SnO$_{2}$, respectively. The $1/\tau_{\mathbf{q},s}^{iso}$ is inversely proportional to atomic mass~\cite{li2014shengbte} which is not very small for tin oxides.

\begin{figure}[tbp!]
\centerline{\includegraphics[width=0.45\textwidth]{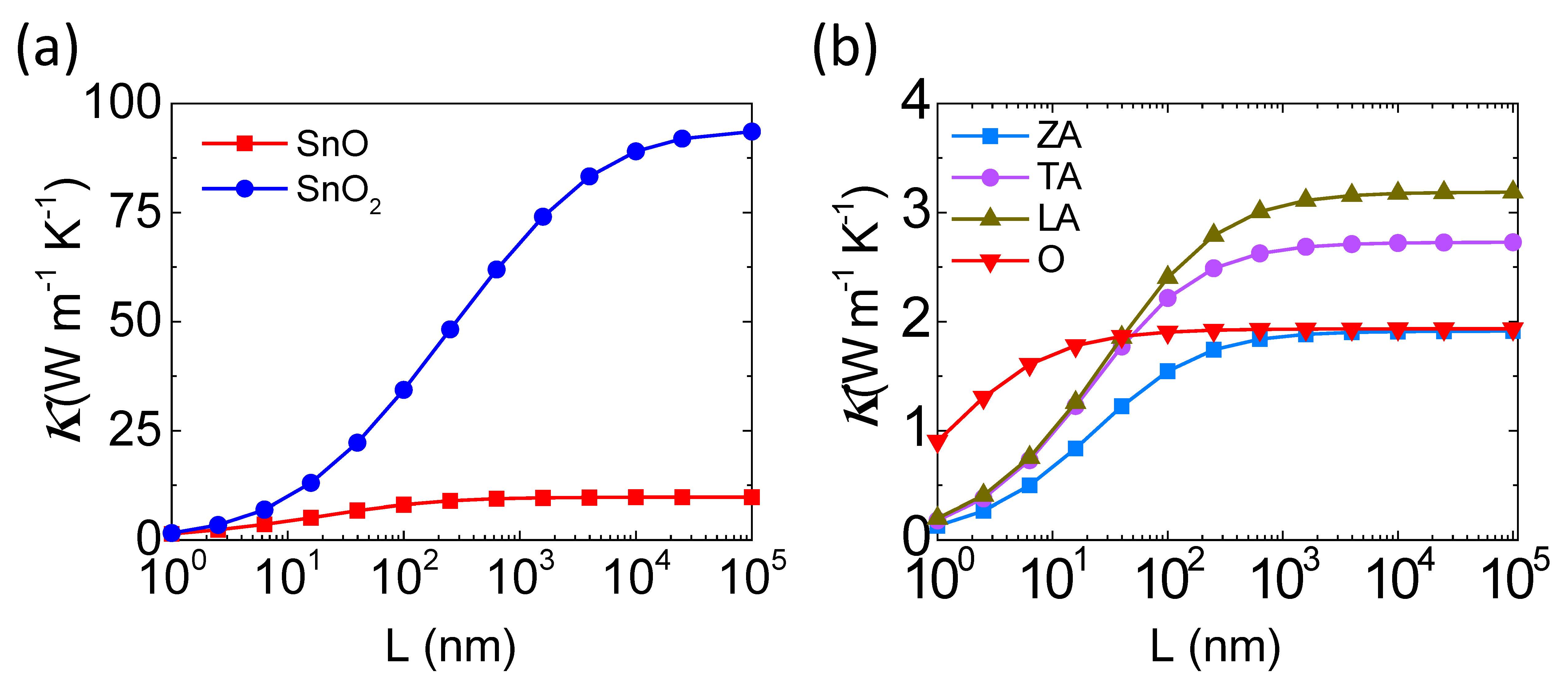}}
\caption{(a) The lateral size ($L$) dependence of the $\kappa$ of monolayer SnO and SnO$_{2}$. (b) The $L$ dependence of the $\kappa$ contributed by different branches in monolayer SnO.}
\label{wh3}
\end{figure}

\begin{figure}[tbp!]
\centerline{\includegraphics[width=0.45\textwidth]{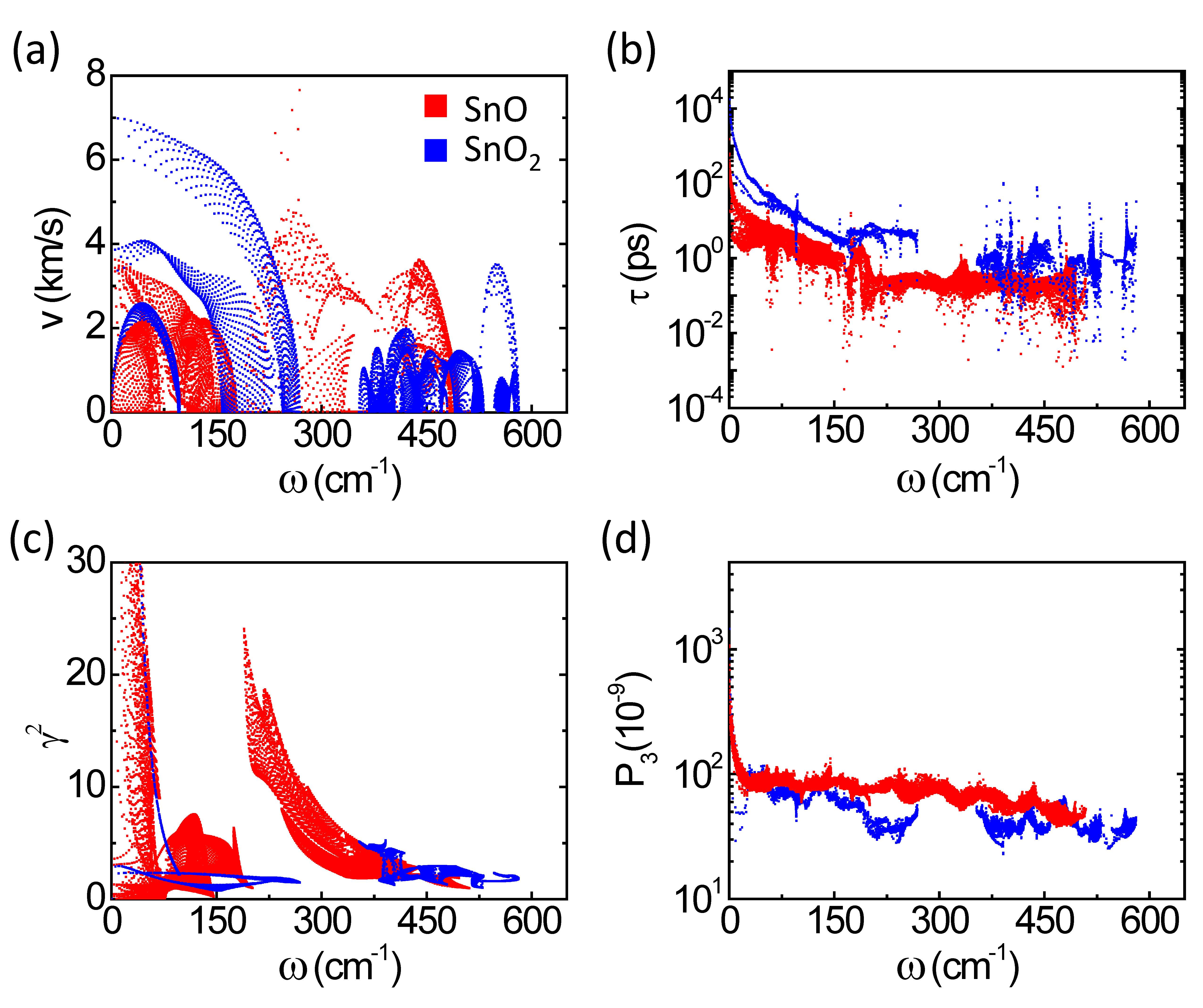}}
\caption{At room temperature, (a) mode group velocity $v_{\mathbf{q},s}$, (b) mode phonon life time $\tau_{\mathbf{q},s}$, (c) The square of mode gr\"{u}neisen parameter $\gamma^{2}_{\mathbf{q},s}$ and (d) phase space for anharmonic scattering $P_{3\mathbf{q},s}$ for monolayer SnO (red) and SnO$_{2}$ (blue), respectively.}
\label{wh4}
\end{figure}

The normalized cumulative $\kappa(\omega < \omega _{0})$, which represents the contribution of $\kappa$ from the phonons with frequencies less than $\omega _{0}$, is shown in Fig.~\ref{wh2}(d).
In the case of monolayer SnO, apart from the acoustic phonons, optical phonons with frequency up to 470 cm$^{-1}$ have a non-negligible contribution ($\sim$ 20\%) to $\kappa$ (see Fig.~\ref{wh2}(d) and Fig. S8(a) in the supplementary material). That can be attributed to the high dispersive relation and large group velocity of optical phonons (see Fig.~\ref{wh2}(a) and Fig.~\ref{wh4}(a)). In contrast, the $\kappa$ of monolayer SnO$_{2}$ is dominated by acoustic phonons with frequency less than 250 cm$^{-1}$ (see Fig.~\ref{wh2}(d)).

The phonon scattering from rough edge can decrease the $\kappa$ of semiconductors to achieve
excellent thermoelectricity.~\cite{silinano1}
Here the size effect of $\kappa$ is discussed in the range of diffusive thermal transport.
The lateral size ($L$) dependence of room-temperature $\kappa$ and the contribution of phonons from different branches are displayed in Fig.~\ref{wh3}. At room temperature, the mean free path (MFP) of phonon is about $10^{3}$ and $10^{5}$ nm for monolayer SnO and SnO$_{2}$, respectively. In the case of monolayer SnO, though the contribution of $\kappa$ from acoustic phonons can be decreased by enhanced boundary
scattering, the contribution from optical phonons is almost size independent until the $L$ decreases down to 10 nm (see Fig.~\ref{wh3}(b)), due to its small MFP. Thus, as the lateral size of 2D SnO decreases from $10^{5}$ nm, nanostructuring might not be an efficient method to reduce its $\kappa$, due to the coexistence of size dependent and independent component. In contrast, the $\kappa$ of 2D SnO$_{2}$, which is dominated by acoustic phonons, can be effectively decreased by edge roughness scattering (see Fig.~\ref{wh3}(a)).

The large difference between the $\kappa$ of monolayer SnO and SnO$_{2}$ can not be simply attributed to the different mean atomic mass. We analyzed the role of each term in Eq.~\ref{kappa} in determining the $\kappa$ of 2D tin oxides.
Figure~\ref{wh4}(a) displays the frequency dependence of mode group velocity $v_{\mathbf{q},s}$. The $v_{\mathbf{q},s}$ of low-frequency phonon modes ($\omega < 170$ cm$^{-1}$) in monolayer SnO$_{2}$ is higher than that in monolayer SnO.
The group velocity of phonon is large for materials with light atomic mass and strong bonding. The bond strength can be reflected by the 2D elastic module $C_{\rm 2D}$, which can be obtained through a fitting process of total energy with respect to strain.~\cite{qiao2014high} The $C_{\rm 2D}$ of monolayer SnO along $x$- and $y$-axis is 40.96 J/m$^{2}$ and 34.98 J/m$^{2}$, much smaller than $C_{\rm 2D}=260.53$ J/m$^{2}$ of monolayer SnO$_{2}$ (see Fig. S9 in the supplementary material). The weak Sn-O bonding results in small phonon group velocity as well as aforementioned long Sn-O bond length in monolayer SnO (see Table.~\ref{table1}).

The mode phonon lifetimes $\tau_{\mathbf{q},s}$ of 2D tin oxides at room temperature are shown in Fig.~\ref{wh4}(b). It can be seen that the overall $\tau_{\mathbf{q},s}$ of monolayer SnO$_{2}$ is larger than that of monolayer SnO.
As mentioned above, anharmonic effect dominates the phonon scattering in the thermal transport of 2D tin oxides. The scattering rate $1/\tau_{\mathbf{q},s}^{an}$ depends on both the square of Gr\"{u}neisen parameter $\gamma_{\mathbf{q},s}^{2}$ and the available three-phonon phase space (P$_{3}$)$_{\mathbf{q},s}$.~\cite{ziman} The former one describes the strength of anharmonicity. The latter one represents the number of anharmonic scattering channel for phonon absorption processes and emission processes, according to the energy and momentum conservations.~\cite{li2014shengbte}

The overall $\gamma^{2}_{\mathbf{q},s}$ of monolayer SnO is much larger than that of monolayer SnO$_{2}$, which can be seen from Fig.~\ref{wh4}(c). That is also consistent with the fact that the acoustic phonon branches of monolayer SnO are significantly softened compared to that of monolayer SnO$_{2}$ (see Fig.~\ref{wh2}). The overall $(P_{3})_{\mathbf{q}, s}$ of monolayer SnO is slightly larger than that of SnO$_{2}$ (see Fig.~\ref{wh4}(d)).
In the phonon dispersion of monolayer SnO, the soft acoustic phonons and intersection between acoustic and optical branches (see Fig.~\ref{wh2}(a)) provide more channels for anharmonic scattering, especially for emission process of high-frequency phonons (see Fig. S10 in the supplementary material). Combining with Fig.~\ref{wh4}(c) and ~\ref{wh4}(d), the small $\tau_{\mathbf{q},s}$ in monolayer SnO is mainly due to the large $\gamma^{2}_{\mathbf{q},s}$. Thus, the much small $\kappa$ of monolayer SnO arises from the small phonon group velocity and strong anharmonicity strength.

\begin{figure}[tbp!]
\centerline{\includegraphics[width=0.45\textwidth]{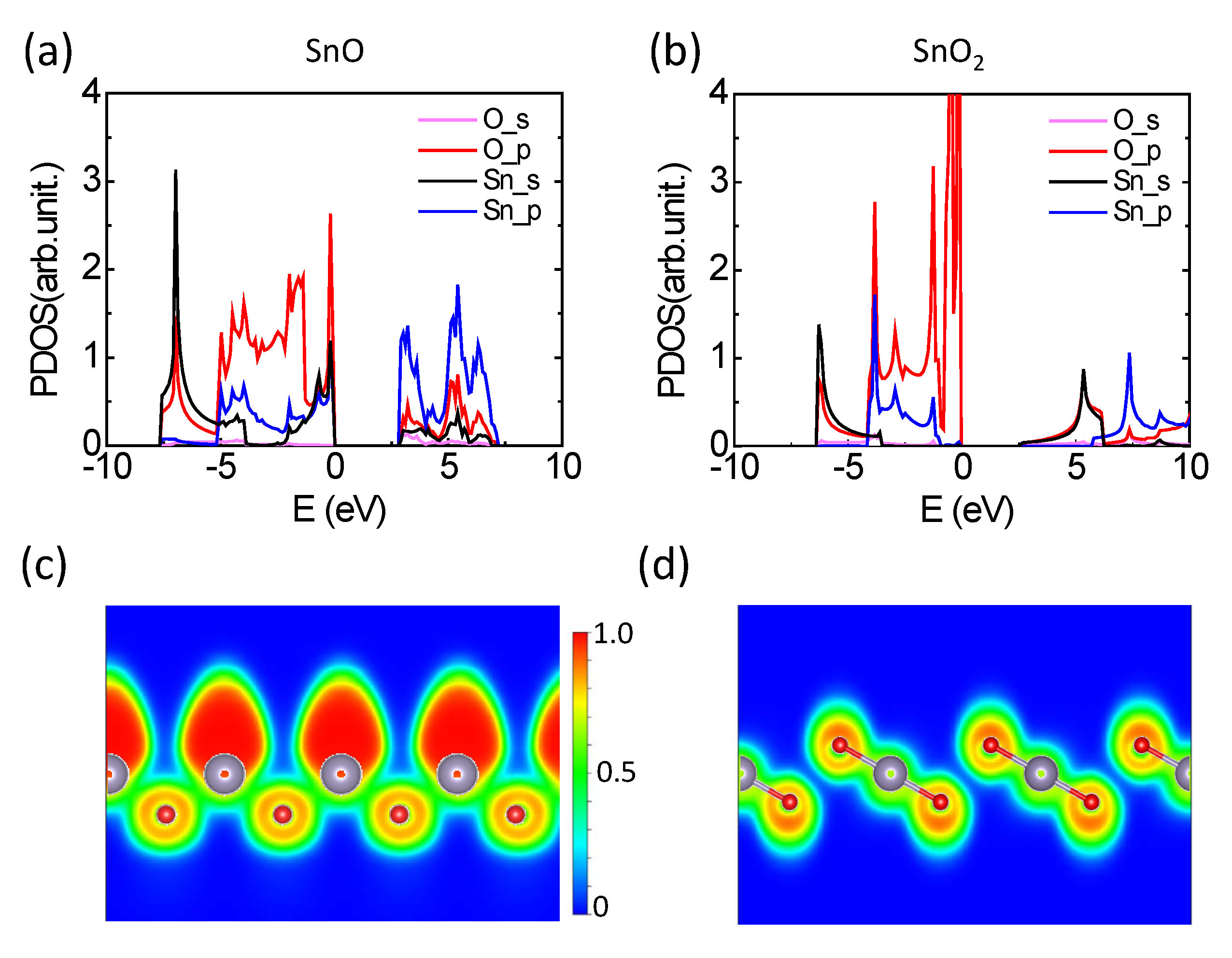}}
\caption{(a, b) are projected density of states of monolayer (a) SnO and (b) SnO$_{2}$, respectively. The electronic local function of monolayer (c) SnO and (d) SnO$_{2}$.}
\label{wh5}
\end{figure}

The large imparity between the anharmonicity strength of tin oxides can be further attributed to the electron configuration.
The projected density of states (PDOS) of tin oxides are shown in Fig.~\ref{wh5}(a) and~\ref{wh5}(b). In the valence band of monolayer SnO, Sn-5$s$ electrons interact with the O-2$p$ states, with the bonding states locating between -8 and -5 eV and anti-bonding states appears at upper valence bands (see Fig.~\ref{wh5}(a) and Fig. S11 in the supplementary material).
The anti-bonding states, with a large contribution from the Sn-5$s$ states, can hybridize with Sn-5$p$ states due to the lack of inversion symmetry. This sort of hybridization shifts the energy of anti-bonding states downward, thereby stabilizing it at the top of valence bands.\cite{bulkPbO} The filled anti-bonding states result in the sterically active lone-pair electrons (see Fig. S11 in the supplementary material), similar to the case of bulk SnO.~\cite{bulkSnO}
The electronic local function (ELF) was calculated with the location of cross-section indicated by the blue line in Fig.~\ref{wh1}(a, b), which can directly give a picture of lone-pair electrons. The ELF shows a strong electron localization around the Sn atom (see Fig.~\ref{wh5}(c)), consistent with the distribution of lone pair at Sn.
As a result, the overlapping wave functions of the lone-pair electrons and nearby O-2$p$ bonding electrons will induce a nonlinear repulsive electrostatic force during thermal agitation,~\cite{lonepair,qin2018} leading to enhanced anharmonicity and small phonon lifetime.
Additionally, it is indicated that chemical functionalization can effectively modulate the $\kappa$ of 2D SnO since the lone-pair electrons distribute on the surface.

In contrast to SnO, the lattice crystal of monolayer SnO$_{2}$ is centrosymmetric. The mixing between Sn-$5p$ states and anti-bonding states of O-$2p$ and Sn-$5s$  is prohibited due to the different parity of $s$ and $p$ orbital. The anti-bonding states of O-$2p$ and Sn-$5s$ is above the Fermi level and unfilled. Thus, both the $s$ and $p$ electrons of Sn atom participate in Sn-O bonding in a form of $sp^{3}$ hybridization. Electrons transfer from Sn to O atom, as shown in Fig.~\ref{wh5}(d) and Fig. S11.
The absence of lone-pair electrons make anharmonicity strength to be smaller for monolayer SnO$_{2}$ than for monolayer SnO (see Fig.~\ref{wh4}(c)), resulting in longer phonon life time (see Fig.~\ref{wh4}(b)).

In conclusion, based on the first-principles calculations, we calculated the thermal conductivity of monolayer SnO and SnO$_{2}$. The tetragonal lattice of monolayer SnO spontaneously transforms to a rectangle one.
The room-temperature $\kappa$ of monolayer SnO and SnO$_{2}$ arrives at 9.6 W/(m$\cdot$K) and 98.8 W/(m$\cdot$K), respectively.
The size effect of $\kappa$ is much weaker for monolayer SnO than for monolayer SnO$_{2}$, due to the coexistence of size-dependent and size-independent contribution of $\kappa$ in monolayer SnO. The small $\kappa$ of monolayer SnO is due to the small phonon group velocity and strong anharmonicity strength. That can be further attributed to the existence of sterically active lone-pair electrons distributed on the surface of monolayer SnO.
This work may give a guide for the thermal management in the electronic or thermoelectrical devices based on 2D tin oxides.

\section*{Supplementary Material}
{{See supplementary material for the computational details; The convergence test of $\kappa$; The phonon dispersion of monolayer, bilayer and bulk SnO with a = b; The structural and energetical properties of bilayer SnO$_{2}$; The 2D effective elastic module, band structure, projected phonon density of states of monolayer SnO and SnO$_{2}$; The isotropic $\kappa$ of monolayer SnO; The temperature dependence of $\kappa$ of monolayer SnO and SnO$_{2}$; The contribution of $\kappa$ from different branches; The three-phonon scattering phase space $(P_{3})_{\mathbf{q}, s}$ for absorption and emission process; The distribution of charge density of monolayer SnO and SnO$_{2}$.}}

\begin{acknowledgments}

This work was supported by the NSFC (Grants No.11747054), the Specialized Research Fund for the
Doctoral Program of Higher Education of China (Grant No.2018M631760), the Project of Heibei Educational Department, China (No. ZD2018015 and QN2018012), and the Advanced Postdoctoral Programs of Hebei Province (No.B2017003004).
\end{acknowledgments}

\nocite{*}
\bibliography{thsno}

\begin{thebibliography}{45}%
\makeatletter
\providecommand \@ifxundefined [1]{%
 \@ifx{#1\undefined}
}%
\providecommand \@ifnum [1]{%
 \ifnum #1\expandafter \@firstoftwo
 \else \expandafter \@secondoftwo
 \fi
}%
\providecommand \@ifx [1]{%
 \ifx #1\expandafter \@firstoftwo
 \else \expandafter \@secondoftwo
 \fi
}%
\providecommand \natexlab [1]{#1}%
\providecommand \enquote  [1]{``#1''}%
\providecommand \bibnamefont  [1]{#1}%
\providecommand \bibfnamefont [1]{#1}%
\providecommand \citenamefont [1]{#1}%
\providecommand \href@noop [0]{\@secondoftwo}%
\providecommand \href [0]{\begingroup \@sanitize@url \@href}%
\providecommand \@href[1]{\@@startlink{#1}\@@href}%
\providecommand \@@href[1]{\endgroup#1\@@endlink}%
\providecommand \@sanitize@url [0]{\catcode `\\12\catcode `\$12\catcode
  `\&12\catcode `\#12\catcode `\^12\catcode `\_12\catcode `\%12\relax}%
\providecommand \@@startlink[1]{}%
\providecommand \@@endlink[0]{}%
\providecommand \url  [0]{\begingroup\@sanitize@url \@url }%
\providecommand \@url [1]{\endgroup\@href {#1}{\urlprefix }}%
\providecommand \urlprefix  [0]{URL }%
\providecommand \Eprint [0]{\href }%
\providecommand \doibase [0]{http://dx.doi.org/}%
\providecommand \selectlanguage [0]{\@gobble}%
\providecommand \bibinfo  [0]{\@secondoftwo}%
\providecommand \bibfield  [0]{\@secondoftwo}%
\providecommand \translation [1]{[#1]}%
\providecommand \BibitemOpen [0]{}%
\providecommand \bibitemStop [0]{}%
\providecommand \bibitemNoStop [0]{.\EOS\space}%
\providecommand \EOS [0]{\spacefactor3000\relax}%
\providecommand \BibitemShut  [1]{\csname bibitem#1\endcsname}%
\let\auto@bib@innerbib\@empty
\bibitem [{\citenamefont {Novoselov}\ \emph {et~al.}(2005)\citenamefont
  {Novoselov}, \citenamefont {Geim}, \citenamefont {Morozov}, \citenamefont
  {Jiang}, \citenamefont {Katsnelson}, \citenamefont {Grigorieva},
  \citenamefont {Dubonos},\ and\ \citenamefont {Firsov}}]{Novoselov2005}%
  \BibitemOpen
  \bibfield  {author} {\bibinfo {author} {\bibfnamefont {K.~S.}\ \bibnamefont
  {Novoselov}}, \bibinfo {author} {\bibfnamefont {A.~K.}\ \bibnamefont {Geim}},
  \bibinfo {author} {\bibfnamefont {S.~V.}\ \bibnamefont {Morozov}}, \bibinfo
  {author} {\bibfnamefont {D.}~\bibnamefont {Jiang}}, \bibinfo {author}
  {\bibfnamefont {M.~I.}\ \bibnamefont {Katsnelson}}, \bibinfo {author}
  {\bibfnamefont {I.~V.}\ \bibnamefont {Grigorieva}}, \bibinfo {author}
  {\bibfnamefont {S.~V.}\ \bibnamefont {Dubonos}}, \ and\ \bibinfo {author}
  {\bibfnamefont {A.~A.}\ \bibnamefont {Firsov}},\ }\href
  {http://dx.doi.org/10.1038/nature04233} {\bibfield  {journal} {\bibinfo
  {journal} {Nature}\ }\textbf {\bibinfo {volume} {438}},\ \bibinfo {pages}
  {197} (\bibinfo {year} {2005})}\BibitemShut {NoStop}%
\bibitem [{\citenamefont {Chhowalla}\ \emph {et~al.}(2013)\citenamefont
  {Chhowalla}, \citenamefont {Shin}, \citenamefont {Eda}, \citenamefont {Li},
  \citenamefont {Loh},\ and\ \citenamefont {Zhang}}]{chhowalla2013}%
  \BibitemOpen
  \bibfield  {author} {\bibinfo {author} {\bibfnamefont {M.}~\bibnamefont
  {Chhowalla}}, \bibinfo {author} {\bibfnamefont {H.~S.}\ \bibnamefont {Shin}},
  \bibinfo {author} {\bibfnamefont {G.}~\bibnamefont {Eda}}, \bibinfo {author}
  {\bibfnamefont {L.-J.}\ \bibnamefont {Li}}, \bibinfo {author} {\bibfnamefont
  {K.~P.}\ \bibnamefont {Loh}}, \ and\ \bibinfo {author} {\bibfnamefont
  {H.}~\bibnamefont {Zhang}},\ }\href {http://dx.doi.org/10.1038/nchem.1589}
  {\bibfield  {journal} {\bibinfo  {journal} {Nat. Chem.}\ }\textbf {\bibinfo
  {volume} {5}},\ \bibinfo {pages} {263} (\bibinfo {year} {2013})}\BibitemShut
  {NoStop}%
\bibitem [{\citenamefont {Nicolosi}\ \emph {et~al.}(2013)\citenamefont
  {Nicolosi}, \citenamefont {Chhowalla}, \citenamefont {Kanatzidis},
  \citenamefont {Strano},\ and\ \citenamefont {Coleman}}]{Nicolosi1226419}%
  \BibitemOpen
  \bibfield  {author} {\bibinfo {author} {\bibfnamefont {V.}~\bibnamefont
  {Nicolosi}}, \bibinfo {author} {\bibfnamefont {M.}~\bibnamefont {Chhowalla}},
  \bibinfo {author} {\bibfnamefont {M.~G.}\ \bibnamefont {Kanatzidis}},
  \bibinfo {author} {\bibfnamefont {M.~S.}\ \bibnamefont {Strano}}, \ and\
  \bibinfo {author} {\bibfnamefont {J.~N.}\ \bibnamefont {Coleman}},\ }\href
  {\doibase 10.1126/science.1226419} {\bibfield  {journal} {\bibinfo  {journal}
  {Science}\ }\textbf {\bibinfo {volume} {340}},\ \bibinfo {pages} {1226419}
  (\bibinfo {year} {2013})}\BibitemShut {NoStop}%
\bibitem [{\citenamefont {Ziletti}\ \emph {et~al.}(2015)\citenamefont
  {Ziletti}, \citenamefont {Carvalho}, \citenamefont {Campbell}, \citenamefont
  {Coker},\ and\ \citenamefont {Castro~Neto}}]{Phosphorene1}%
  \BibitemOpen
  \bibfield  {author} {\bibinfo {author} {\bibfnamefont {A.}~\bibnamefont
  {Ziletti}}, \bibinfo {author} {\bibfnamefont {A.}~\bibnamefont {Carvalho}},
  \bibinfo {author} {\bibfnamefont {D.~K.}\ \bibnamefont {Campbell}}, \bibinfo
  {author} {\bibfnamefont {D.~F.}\ \bibnamefont {Coker}}, \ and\ \bibinfo
  {author} {\bibfnamefont {A.~H.}\ \bibnamefont {Castro~Neto}},\ }\href
  {\doibase 10.1103/PhysRevLett.114.046801} {\bibfield  {journal} {\bibinfo
  {journal} {Phys. Rev. Lett.}\ }\textbf {\bibinfo {volume} {114}},\ \bibinfo
  {pages} {046801} (\bibinfo {year} {2015})}\BibitemShut {NoStop}%
\bibitem [{\citenamefont {Cui}\ \emph {et~al.}(2015)\citenamefont {Cui},
  \citenamefont {Wen}, \citenamefont {Huang}, \citenamefont {Chang},\ and\
  \citenamefont {Chen}}]{MoS2O}%
  \BibitemOpen
  \bibfield  {author} {\bibinfo {author} {\bibfnamefont {S.}~\bibnamefont
  {Cui}}, \bibinfo {author} {\bibfnamefont {Z.}~\bibnamefont {Wen}}, \bibinfo
  {author} {\bibfnamefont {X.}~\bibnamefont {Huang}}, \bibinfo {author}
  {\bibfnamefont {J.}~\bibnamefont {Chang}}, \ and\ \bibinfo {author}
  {\bibfnamefont {J.}~\bibnamefont {Chen}},\ }\href {\doibase
  10.1002/smll.201402923} {\bibfield  {journal} {\bibinfo  {journal} {Small}\
  }\textbf {\bibinfo {volume} {11}},\ \bibinfo {pages} {2305--2313} (\bibinfo
  {year} {2015})}\BibitemShut {NoStop}%
\bibitem [{\citenamefont {Daeneke}\ \emph {et~al.}(2017)\citenamefont
  {Daeneke}, \citenamefont {Atkin}, \citenamefont {Orrell-Trigg}, \citenamefont
  {Zavabeti}, \citenamefont {Ahmed}, \citenamefont {Walia}, \citenamefont
  {Liu}, \citenamefont {Tachibana}, \citenamefont {Javaid}, \citenamefont
  {Greentree}, \citenamefont {Russo}, \citenamefont {Kaner},\ and\
  \citenamefont {Kalantar-Zadeh}}]{SnOproduce1}%
  \BibitemOpen
  \bibfield  {author} {\bibinfo {author} {\bibfnamefont {T.}~\bibnamefont
  {Daeneke}}, \bibinfo {author} {\bibfnamefont {P.}~\bibnamefont {Atkin}},
  \bibinfo {author} {\bibfnamefont {R.}~\bibnamefont {Orrell-Trigg}}, \bibinfo
  {author} {\bibfnamefont {A.}~\bibnamefont {Zavabeti}}, \bibinfo {author}
  {\bibfnamefont {T.}~\bibnamefont {Ahmed}}, \bibinfo {author} {\bibfnamefont
  {S.}~\bibnamefont {Walia}}, \bibinfo {author} {\bibfnamefont
  {M.}~\bibnamefont {Liu}}, \bibinfo {author} {\bibfnamefont {Y.}~\bibnamefont
  {Tachibana}}, \bibinfo {author} {\bibfnamefont {M.}~\bibnamefont {Javaid}},
  \bibinfo {author} {\bibfnamefont {A.~D.}\ \bibnamefont {Greentree}}, \bibinfo
  {author} {\bibfnamefont {S.~P.}\ \bibnamefont {Russo}}, \bibinfo {author}
  {\bibfnamefont {R.~B.}\ \bibnamefont {Kaner}}, \ and\ \bibinfo {author}
  {\bibfnamefont {K.}~\bibnamefont {Kalantar-Zadeh}},\ }\href {\doibase
  10.1021/acsnano.7b04856} {\bibfield  {journal} {\bibinfo  {journal} {ACS
  Nano}\ }\textbf {\bibinfo {volume} {11}},\ \bibinfo {pages} {10974--10983}
  (\bibinfo {year} {2017})}\BibitemShut {NoStop}%
\bibitem [{\citenamefont {Sui}\ and\ \citenamefont
  {Charpentier}(2012)}]{synthesis}%
  \BibitemOpen
  \bibfield  {author} {\bibinfo {author} {\bibfnamefont {R.}~\bibnamefont
  {Sui}}\ and\ \bibinfo {author} {\bibfnamefont {P.}~\bibnamefont
  {Charpentier}},\ }\href {\doibase 10.1021/cr2000465} {\bibfield  {journal}
  {\bibinfo  {journal} {Chem. Rev.}\ }\textbf {\bibinfo {volume} {112}},\
  \bibinfo {pages} {3057--3082} (\bibinfo {year} {2012})}\BibitemShut {NoStop}%
\bibitem [{\citenamefont {Tokura}\ and\ \citenamefont
  {Nagaosa}(2000)}]{Tokura462}%
  \BibitemOpen
  \bibfield  {author} {\bibinfo {author} {\bibfnamefont {Y.}~\bibnamefont
  {Tokura}}\ and\ \bibinfo {author} {\bibfnamefont {N.}~\bibnamefont
  {Nagaosa}},\ }\href {\doibase 10.1126/science.288.5465.462} {\bibfield
  {journal} {\bibinfo  {journal} {Science}\ }\textbf {\bibinfo {volume}
  {288}},\ \bibinfo {pages} {462--468} (\bibinfo {year} {2000})}\BibitemShut
  {NoStop}%
\bibitem [{\citenamefont {Miller}\ \emph {et~al.}(2017)\citenamefont {Miller},
  \citenamefont {Gorai}, \citenamefont {Aydemir}, \citenamefont {Mason},
  \citenamefont {Stevanović}, \citenamefont {Toberer},\ and\ \citenamefont
  {Snyder}}]{SnOthermal}%
  \BibitemOpen
  \bibfield  {author} {\bibinfo {author} {\bibfnamefont {S.~A.}\ \bibnamefont
  {Miller}}, \bibinfo {author} {\bibfnamefont {P.}~\bibnamefont {Gorai}},
  \bibinfo {author} {\bibfnamefont {U.}~\bibnamefont {Aydemir}}, \bibinfo
  {author} {\bibfnamefont {T.~O.}\ \bibnamefont {Mason}}, \bibinfo {author}
  {\bibfnamefont {V.}~\bibnamefont {Stevanović}}, \bibinfo {author}
  {\bibfnamefont {E.~S.}\ \bibnamefont {Toberer}}, \ and\ \bibinfo {author}
  {\bibfnamefont {G.~J.}\ \bibnamefont {Snyder}},\ }\href {\doibase
  10.1039/C7TC01623A} {\bibfield  {journal} {\bibinfo  {journal} {J. Mater.
  Chem. C}\ }\textbf {\bibinfo {volume} {5}},\ \bibinfo {pages} {8854--8861}
  (\bibinfo {year} {2017})}\BibitemShut {NoStop}%
\bibitem [{\citenamefont {Padova}\ \emph {et~al.}(1994)\citenamefont {Padova},
  \citenamefont {Fanfoni}, \citenamefont {Larciprete}, \citenamefont
  {Mangiantini}, \citenamefont {Priori},\ and\ \citenamefont
  {Perfetti}}]{DEPADOVA1994379}%
  \BibitemOpen
  \bibfield  {author} {\bibinfo {author} {\bibfnamefont {P.~D.}\ \bibnamefont
  {Padova}}, \bibinfo {author} {\bibfnamefont {M.}~\bibnamefont {Fanfoni}},
  \bibinfo {author} {\bibfnamefont {R.}~\bibnamefont {Larciprete}}, \bibinfo
  {author} {\bibfnamefont {M.}~\bibnamefont {Mangiantini}}, \bibinfo {author}
  {\bibfnamefont {S.}~\bibnamefont {Priori}}, \ and\ \bibinfo {author}
  {\bibfnamefont {P.}~\bibnamefont {Perfetti}},\ }\href {\doibase
  https://doi.org/10.1016/0039-6028(94)90058-2} {\bibfield  {journal} {\bibinfo
   {journal} {Surf. Sci.}\ }\textbf {\bibinfo {volume} {313}},\ \bibinfo
  {pages} {379 -- 391} (\bibinfo {year} {1994})}\BibitemShut {NoStop}%
\bibitem [{\citenamefont {Lee}\ \emph {et~al.}(2018)\citenamefont {Lee},
  \citenamefont {Yoo}, \citenamefont {Kang}, \citenamefont {Lee}, \citenamefont
  {Choi}, \citenamefont {Park}, \citenamefont {Yi}, \citenamefont {Kim},\ and\
  \citenamefont {Park}}]{SnOSnO2}%
  \BibitemOpen
  \bibfield  {author} {\bibinfo {author} {\bibfnamefont {J.-H.}\ \bibnamefont
  {Lee}}, \bibinfo {author} {\bibfnamefont {M.}~\bibnamefont {Yoo}}, \bibinfo
  {author} {\bibfnamefont {D.}~\bibnamefont {Kang}}, \bibinfo {author}
  {\bibfnamefont {H.-M.}\ \bibnamefont {Lee}}, \bibinfo {author} {\bibfnamefont
  {W.-h.}\ \bibnamefont {Choi}}, \bibinfo {author} {\bibfnamefont {J.~W.}\
  \bibnamefont {Park}}, \bibinfo {author} {\bibfnamefont {Y.}~\bibnamefont
  {Yi}}, \bibinfo {author} {\bibfnamefont {H.~Y.}\ \bibnamefont {Kim}}, \ and\
  \bibinfo {author} {\bibfnamefont {J.-S.}\ \bibnamefont {Park}},\ }\href
  {\doibase 10.1021/acsami.8b12251} {\bibfield  {journal} {\bibinfo  {journal}
  {ACS Appl. Mater. Interfaces}\ }\textbf {\bibinfo {volume} {10}},\ \bibinfo
  {pages} {33335--33342} (\bibinfo {year} {2018})}\BibitemShut {NoStop}%
\bibitem [{\citenamefont {Hien}\ \emph {et~al.}(2014)\citenamefont {Hien},
  \citenamefont {Lee}, \citenamefont {Kim},\ and\ \citenamefont
  {Heo}}]{HIEN2014134}%
  \BibitemOpen
  \bibfield  {author} {\bibinfo {author} {\bibfnamefont {V.~X.}\ \bibnamefont
  {Hien}}, \bibinfo {author} {\bibfnamefont {J.-H.}\ \bibnamefont {Lee}},
  \bibinfo {author} {\bibfnamefont {J.-J.}\ \bibnamefont {Kim}}, \ and\
  \bibinfo {author} {\bibfnamefont {Y.-W.}\ \bibnamefont {Heo}},\ }\href
  {\doibase https://doi.org/10.1016/j.snb.2013.12.086} {\bibfield  {journal}
  {\bibinfo  {journal} {Sens. Actuators, B}\ }\textbf {\bibinfo {volume}
  {194}},\ \bibinfo {pages} {134 -- 141} (\bibinfo {year} {2014})}\BibitemShut
  {NoStop}%
\bibitem [{\citenamefont {Kumar}\ \emph {et~al.}(2009)\citenamefont {Kumar},
  \citenamefont {Sen}, \citenamefont {Muthe}, \citenamefont {Gaur},
  \citenamefont {Gupta},\ and\ \citenamefont {Yakhmi}}]{KUMAR2009587}%
  \BibitemOpen
  \bibfield  {author} {\bibinfo {author} {\bibfnamefont {V.}~\bibnamefont
  {Kumar}}, \bibinfo {author} {\bibfnamefont {S.}~\bibnamefont {Sen}}, \bibinfo
  {author} {\bibfnamefont {K.}~\bibnamefont {Muthe}}, \bibinfo {author}
  {\bibfnamefont {N.}~\bibnamefont {Gaur}}, \bibinfo {author} {\bibfnamefont
  {S.}~\bibnamefont {Gupta}}, \ and\ \bibinfo {author} {\bibfnamefont
  {J.}~\bibnamefont {Yakhmi}},\ }\href {\doibase
  https://doi.org/10.1016/j.snb.2009.02.053} {\bibfield  {journal} {\bibinfo
  {journal} {Sens. Actuators, B}\ }\textbf {\bibinfo {volume} {138}},\ \bibinfo
  {pages} {587 -- 590} (\bibinfo {year} {2009})}\BibitemShut {NoStop}%
\bibitem [{\citenamefont {Qiang}\ \emph {et~al.}(2017)\citenamefont {Qiang},
  \citenamefont {Liu}, \citenamefont {Pei}, \citenamefont {Wang},\ and\
  \citenamefont {Yao}}]{SnOFET}%
  \BibitemOpen
  \bibfield  {author} {\bibinfo {author} {\bibfnamefont {L.}~\bibnamefont
  {Qiang}}, \bibinfo {author} {\bibfnamefont {W.}~\bibnamefont {Liu}}, \bibinfo
  {author} {\bibfnamefont {Y.}~\bibnamefont {Pei}}, \bibinfo {author}
  {\bibfnamefont {G.}~\bibnamefont {Wang}}, \ and\ \bibinfo {author}
  {\bibfnamefont {R.}~\bibnamefont {Yao}},\ }\href {\doibase
  https://doi.org/10.1016/j.sse.2016.11.010} {\bibfield  {journal} {\bibinfo
  {journal} {Solid-State Electron.}\ }\textbf {\bibinfo {volume} {129}},\
  \bibinfo {pages} {163 -- 167} (\bibinfo {year} {2017})}\BibitemShut {NoStop}%
\bibitem [{\citenamefont {Dattoli}\ \emph {et~al.}(2007)\citenamefont
  {Dattoli}, \citenamefont {Wan}, \citenamefont {Guo}, \citenamefont {Chen},
  \citenamefont {Pan},\ and\ \citenamefont {Lu}}]{SnO2FET}%
  \BibitemOpen
  \bibfield  {author} {\bibinfo {author} {\bibfnamefont {E.~N.}\ \bibnamefont
  {Dattoli}}, \bibinfo {author} {\bibfnamefont {Q.}~\bibnamefont {Wan}},
  \bibinfo {author} {\bibfnamefont {W.}~\bibnamefont {Guo}}, \bibinfo {author}
  {\bibfnamefont {Y.}~\bibnamefont {Chen}}, \bibinfo {author} {\bibfnamefont
  {X.}~\bibnamefont {Pan}}, \ and\ \bibinfo {author} {\bibfnamefont
  {W.}~\bibnamefont {Lu}},\ }\href {\doibase 10.1021/nl0712217} {\bibfield
  {journal} {\bibinfo  {journal} {Nano Lett.}\ }\textbf {\bibinfo {volume}
  {7}},\ \bibinfo {pages} {2463--2469} (\bibinfo {year} {2007})}\BibitemShut
  {NoStop}%
\bibitem [{\citenamefont {Zhang}\ \emph {et~al.}(2017)\citenamefont {Zhang},
  \citenamefont {Zhu}, \citenamefont {Zhang}, \citenamefont
  {Schwingenschlögl},\ and\ \citenamefont {Alshareef}}]{SnOanode}%
  \BibitemOpen
  \bibfield  {author} {\bibinfo {author} {\bibfnamefont {F.}~\bibnamefont
  {Zhang}}, \bibinfo {author} {\bibfnamefont {J.}~\bibnamefont {Zhu}}, \bibinfo
  {author} {\bibfnamefont {D.}~\bibnamefont {Zhang}}, \bibinfo {author}
  {\bibfnamefont {U.}~\bibnamefont {Schwingenschlögl}}, \ and\ \bibinfo
  {author} {\bibfnamefont {H.~N.}\ \bibnamefont {Alshareef}},\ }\href {\doibase
  10.1021/acs.nanolett.6b05280} {\bibfield  {journal} {\bibinfo  {journal}
  {Nano Lett.}\ }\textbf {\bibinfo {volume} {17}},\ \bibinfo {pages}
  {1302--1311} (\bibinfo {year} {2017})}\BibitemShut {NoStop}%
\bibitem [{\citenamefont {Yuan}\ \emph {et~al.}(2006)\citenamefont {Yuan},
  \citenamefont {Guo}, \citenamefont {Konstantinov}, \citenamefont {Liu},\ and\
  \citenamefont {Dou}}]{SnO2anode}%
  \BibitemOpen
  \bibfield  {author} {\bibinfo {author} {\bibfnamefont {L.}~\bibnamefont
  {Yuan}}, \bibinfo {author} {\bibfnamefont {Z.}~\bibnamefont {Guo}}, \bibinfo
  {author} {\bibfnamefont {K.}~\bibnamefont {Konstantinov}}, \bibinfo {author}
  {\bibfnamefont {H.}~\bibnamefont {Liu}}, \ and\ \bibinfo {author}
  {\bibfnamefont {S.}~\bibnamefont {Dou}},\ }\href {\doibase
  https://doi.org/10.1016/j.jpowsour.2006.04.048} {\bibfield  {journal}
  {\bibinfo  {journal} {J. Power Sources}\ }\textbf {\bibinfo {volume} {159}},\
  \bibinfo {pages} {345 -- 348} (\bibinfo {year} {2006})}\BibitemShut {NoStop}%
\bibitem [{\citenamefont {Tsai}\ \emph {et~al.}(2011)\citenamefont {Tsai},
  \citenamefont {Wang}, \citenamefont {Tu}, \citenamefont {Hsu}, \citenamefont
  {Kuo}, \citenamefont {Lin},\ and\ \citenamefont {Ko}}]{Snoele}%
  \BibitemOpen
  \bibfield  {author} {\bibinfo {author} {\bibfnamefont {F.-S.}\ \bibnamefont
  {Tsai}}, \bibinfo {author} {\bibfnamefont {S.-J.}\ \bibnamefont {Wang}},
  \bibinfo {author} {\bibfnamefont {Y.-C.}\ \bibnamefont {Tu}}, \bibinfo
  {author} {\bibfnamefont {Y.-W.}\ \bibnamefont {Hsu}}, \bibinfo {author}
  {\bibfnamefont {C.-Y.}\ \bibnamefont {Kuo}}, \bibinfo {author} {\bibfnamefont
  {Z.-S.}\ \bibnamefont {Lin}}, \ and\ \bibinfo {author} {\bibfnamefont
  {R.-M.}\ \bibnamefont {Ko}},\ }\href
  {http://stacks.iop.org/1882-0786/4/i=2/a=025002} {\bibfield  {journal}
  {\bibinfo  {journal} {Appl. Phys Express}\ }\textbf {\bibinfo {volume} {4}},\
  \bibinfo {pages} {025002} (\bibinfo {year} {2011})}\BibitemShut {NoStop}%
\bibitem [{\citenamefont {Nomura}, \citenamefont {Kamiya},\ and\ \citenamefont
  {Hosono}(2011)}]{Nomura2011}%
  \BibitemOpen
  \bibfield  {author} {\bibinfo {author} {\bibfnamefont {K.}~\bibnamefont
  {Nomura}}, \bibinfo {author} {\bibfnamefont {T.}~\bibnamefont {Kamiya}}, \
  and\ \bibinfo {author} {\bibfnamefont {H.}~\bibnamefont {Hosono}},\ }\href
  {\doibase 10.1002/adma.201101410} {\bibfield  {journal} {\bibinfo  {journal}
  {Adv. Mater.}\ }\textbf {\bibinfo {volume} {23}},\ \bibinfo {pages}
  {3431--3434} (\bibinfo {year} {2011})}\BibitemShut {NoStop}%
\bibitem [{\citenamefont {Ogale}\ \emph {et~al.}(2003)\citenamefont {Ogale},
  \citenamefont {Choudhary}, \citenamefont {Buban}, \citenamefont {Lofland},
  \citenamefont {Shinde}, \citenamefont {Kale}, \citenamefont {Kulkarni},
  \citenamefont {Higgins}, \citenamefont {Lanci}, \citenamefont {Simpson},
  \citenamefont {Browning}, \citenamefont {Das~Sarma}, \citenamefont {Drew},
  \citenamefont {Greene},\ and\ \citenamefont {Venkatesan}}]{SnO2mage}%
  \BibitemOpen
  \bibfield  {author} {\bibinfo {author} {\bibfnamefont {S.~B.}\ \bibnamefont
  {Ogale}}, \bibinfo {author} {\bibfnamefont {R.~J.}\ \bibnamefont
  {Choudhary}}, \bibinfo {author} {\bibfnamefont {J.~P.}\ \bibnamefont
  {Buban}}, \bibinfo {author} {\bibfnamefont {S.~E.}\ \bibnamefont {Lofland}},
  \bibinfo {author} {\bibfnamefont {S.~R.}\ \bibnamefont {Shinde}}, \bibinfo
  {author} {\bibfnamefont {S.~N.}\ \bibnamefont {Kale}}, \bibinfo {author}
  {\bibfnamefont {V.~N.}\ \bibnamefont {Kulkarni}}, \bibinfo {author}
  {\bibfnamefont {J.}~\bibnamefont {Higgins}}, \bibinfo {author} {\bibfnamefont
  {C.}~\bibnamefont {Lanci}}, \bibinfo {author} {\bibfnamefont {J.~R.}\
  \bibnamefont {Simpson}}, \bibinfo {author} {\bibfnamefont {N.~D.}\
  \bibnamefont {Browning}}, \bibinfo {author} {\bibfnamefont {S.}~\bibnamefont
  {Das~Sarma}}, \bibinfo {author} {\bibfnamefont {H.~D.}\ \bibnamefont {Drew}},
  \bibinfo {author} {\bibfnamefont {R.~L.}\ \bibnamefont {Greene}}, \ and\
  \bibinfo {author} {\bibfnamefont {T.}~\bibnamefont {Venkatesan}},\ }\href
  {\doibase 10.1103/PhysRevLett.91.077205} {\bibfield  {journal} {\bibinfo
  {journal} {Phys. Rev. Lett.}\ }\textbf {\bibinfo {volume} {91}},\ \bibinfo
  {pages} {077205} (\bibinfo {year} {2003})}\BibitemShut {NoStop}%
\bibitem [{\citenamefont {Singh}\ \emph {et~al.}(2017)\citenamefont {Singh},
  \citenamefont {Gaspera}, \citenamefont {Ahmed}, \citenamefont {Walia},
  \citenamefont {Ramanathan}, \citenamefont {van Embden}, \citenamefont
  {Mayes},\ and\ \citenamefont {Bansal}}]{SnOoptical}%
  \BibitemOpen
  \bibfield  {author} {\bibinfo {author} {\bibfnamefont {M.}~\bibnamefont
  {Singh}}, \bibinfo {author} {\bibfnamefont {E.~D.}\ \bibnamefont {Gaspera}},
  \bibinfo {author} {\bibfnamefont {T.}~\bibnamefont {Ahmed}}, \bibinfo
  {author} {\bibfnamefont {S.}~\bibnamefont {Walia}}, \bibinfo {author}
  {\bibfnamefont {R.}~\bibnamefont {Ramanathan}}, \bibinfo {author}
  {\bibfnamefont {J.}~\bibnamefont {van Embden}}, \bibinfo {author}
  {\bibfnamefont {E.}~\bibnamefont {Mayes}}, \ and\ \bibinfo {author}
  {\bibfnamefont {V.}~\bibnamefont {Bansal}},\ }\href {http://stacks.iop.orV.
  Nicolosig/2053-1583/4/i=2/a=025110} {\bibfield  {journal} {\bibinfo
  {journal} {2D Materials}\ }\textbf {\bibinfo {volume} {4}},\ \bibinfo {pages}
  {025110} (\bibinfo {year} {2017})}\BibitemShut {NoStop}%
\bibitem [{\citenamefont {Saji}\ \emph {et~al.}(2016)\citenamefont {Saji},
  \citenamefont {Tian}, \citenamefont {Snure},\ and\ \citenamefont
  {Tiwari}}]{SnOproduce}%
  \BibitemOpen
  \bibfield  {author} {\bibinfo {author} {\bibfnamefont {K.~J.}\ \bibnamefont
  {Saji}}, \bibinfo {author} {\bibfnamefont {K.}~\bibnamefont {Tian}}, \bibinfo
  {author} {\bibfnamefont {M.}~\bibnamefont {Snure}}, \ and\ \bibinfo {author}
  {\bibfnamefont {A.}~\bibnamefont {Tiwari}},\ }\href {\doibase
  10.1002/aelm.201500453} {\bibfield  {journal} {\bibinfo  {journal} {Adv.
  Electron. Mater.}\ }\textbf {\bibinfo {volume} {2}},\ \bibinfo {pages}
  {1500453} (\bibinfo {year} {2016})}\BibitemShut {NoStop}%
\bibitem [{\citenamefont {Seixas}\ \emph {et~al.}(2016)\citenamefont {Seixas},
  \citenamefont {Rodin}, \citenamefont {Carvalho},\ and\ \citenamefont
  {Castro~Neto}}]{magneti}%
  \BibitemOpen
  \bibfield  {author} {\bibinfo {author} {\bibfnamefont {L.}~\bibnamefont
  {Seixas}}, \bibinfo {author} {\bibfnamefont {A.~S.}\ \bibnamefont {Rodin}},
  \bibinfo {author} {\bibfnamefont {A.}~\bibnamefont {Carvalho}}, \ and\
  \bibinfo {author} {\bibfnamefont {A.~H.}\ \bibnamefont {Castro~Neto}},\
  }\href {\doibase 10.1103/PhysRevLett.116.206803} {\bibfield  {journal}
  {\bibinfo  {journal} {Phys. Rev. Lett.}\ }\textbf {\bibinfo {volume} {116}},\
  \bibinfo {pages} {206803} (\bibinfo {year} {2016})}\BibitemShut {NoStop}%
\bibitem [{\citenamefont {Chen}\ \emph {et~al.}(2012)\citenamefont {Chen},
  \citenamefont {Hu}, \citenamefont {Fang},\ and\ \citenamefont
  {Wu}}]{SnO2produce}%
  \BibitemOpen
  \bibfield  {author} {\bibinfo {author} {\bibfnamefont {H.}~\bibnamefont
  {Chen}}, \bibinfo {author} {\bibfnamefont {L.}~\bibnamefont {Hu}}, \bibinfo
  {author} {\bibfnamefont {X.}~\bibnamefont {Fang}}, \ and\ \bibinfo {author}
  {\bibfnamefont {L.}~\bibnamefont {Wu}},\ }\href {\doibase
  10.1002/adfm.201102506} {\bibfield  {journal} {\bibinfo  {journal} {Adv.
  Funct. Mater.}\ }\textbf {\bibinfo {volume} {22}},\ \bibinfo {pages}
  {1229--1235} (\bibinfo {year} {2012})}\BibitemShut {NoStop}%
\bibitem [{\citenamefont {Xiao}, \citenamefont {Xiao},\ and\ \citenamefont
  {Wang}(2016)}]{SnO2predict}%
  \BibitemOpen
  \bibfield  {author} {\bibinfo {author} {\bibfnamefont {W.-Z.}\ \bibnamefont
  {Xiao}}, \bibinfo {author} {\bibfnamefont {G.}~\bibnamefont {Xiao}}, \ and\
  \bibinfo {author} {\bibfnamefont {L.-L.}\ \bibnamefont {Wang}},\ }\href
  {\doibase 10.1063/1.4966581} {\bibfield  {journal} {\bibinfo  {journal} {J.
  Chem. Phys.}\ }\textbf {\bibinfo {volume} {145}},\ \bibinfo {pages} {174702}
  (\bibinfo {year} {2016})}\BibitemShut {NoStop}%
\bibitem [{\citenamefont {Luan}\ \emph {et~al.}(2014)\citenamefont {Luan},
  \citenamefont {Zhang}, \citenamefont {Li}, \citenamefont {Li}, \citenamefont
  {Ren}, \citenamefont {Yuan}, \citenamefont {Ji},\ and\ \citenamefont
  {Wang}}]{SnO2mage2}%
  \BibitemOpen
  \bibfield  {author} {\bibinfo {author} {\bibfnamefont {H.-X.}\ \bibnamefont
  {Luan}}, \bibinfo {author} {\bibfnamefont {C.-W.}\ \bibnamefont {Zhang}},
  \bibinfo {author} {\bibfnamefont {F.}~\bibnamefont {Li}}, \bibinfo {author}
  {\bibfnamefont {P.}~\bibnamefont {Li}}, \bibinfo {author} {\bibfnamefont
  {M.-J.}\ \bibnamefont {Ren}}, \bibinfo {author} {\bibfnamefont
  {M.}~\bibnamefont {Yuan}}, \bibinfo {author} {\bibfnamefont {W.-X.}\
  \bibnamefont {Ji}}, \ and\ \bibinfo {author} {\bibfnamefont {P.-J.}\
  \bibnamefont {Wang}},\ }\href {\doibase 10.1039/C3RA46325G} {\bibfield
  {journal} {\bibinfo  {journal} {RSC Adv.}\ }\textbf {\bibinfo {volume} {4}},\
  \bibinfo {pages} {9602--9607} (\bibinfo {year} {2014})}\BibitemShut {NoStop}%
\bibitem [{\citenamefont {Feng}\ \emph {et~al.}(2015)\citenamefont {Feng},
  \citenamefont {Ji}, \citenamefont {Huang}, \citenamefont {Chen},
  \citenamefont {Li}, \citenamefont {Li}, \citenamefont {Zhang},\ and\
  \citenamefont {Wang}}]{SnO2mage3}%
  \BibitemOpen
  \bibfield  {author} {\bibinfo {author} {\bibfnamefont {Y.}~\bibnamefont
  {Feng}}, \bibinfo {author} {\bibfnamefont {W.-X.}\ \bibnamefont {Ji}},
  \bibinfo {author} {\bibfnamefont {B.-J.}\ \bibnamefont {Huang}}, \bibinfo
  {author} {\bibfnamefont {X.-l.}\ \bibnamefont {Chen}}, \bibinfo {author}
  {\bibfnamefont {F.}~\bibnamefont {Li}}, \bibinfo {author} {\bibfnamefont
  {P.}~\bibnamefont {Li}}, \bibinfo {author} {\bibfnamefont {C.-w.}\
  \bibnamefont {Zhang}}, \ and\ \bibinfo {author} {\bibfnamefont {P.-J.}\
  \bibnamefont {Wang}},\ }\href {\doibase 10.1039/C5RA00788G} {\bibfield
  {journal} {\bibinfo  {journal} {RSC Adv.}\ }\textbf {\bibinfo {volume} {5}},\
  \bibinfo {pages} {24306--24312} (\bibinfo {year} {2015})}\BibitemShut
  {NoStop}%
\bibitem [{\citenamefont {Zou}\ and\ \citenamefont {Balandin}(2001)}]{zou2001}%
  \BibitemOpen
  \bibfield  {author} {\bibinfo {author} {\bibfnamefont {J.}~\bibnamefont
  {Zou}}\ and\ \bibinfo {author} {\bibfnamefont {A.}~\bibnamefont {Balandin}},\
  }\href {\doibase 10.1063/1.1345515} {\bibfield  {journal} {\bibinfo
  {journal} {J. Appl. Phys.}\ }\textbf {\bibinfo {volume} {89}},\ \bibinfo
  {pages} {2932--2938} (\bibinfo {year} {2001})}\BibitemShut {NoStop}%
\bibitem [{\citenamefont {Hochbaum}\ \emph {et~al.}(2008)\citenamefont
  {Hochbaum}, \citenamefont {Chen}, \citenamefont {Delgado}, \citenamefont
  {Liang}, \citenamefont {Garnett}, \citenamefont {Najarian}, \citenamefont
  {Majumdar},\ and\ \citenamefont {Yang}}]{silinano1}%
  \BibitemOpen
  \bibfield  {author} {\bibinfo {author} {\bibfnamefont {A.~I.}\ \bibnamefont
  {Hochbaum}}, \bibinfo {author} {\bibfnamefont {R.}~\bibnamefont {Chen}},
  \bibinfo {author} {\bibfnamefont {R.~D.}\ \bibnamefont {Delgado}}, \bibinfo
  {author} {\bibfnamefont {W.}~\bibnamefont {Liang}}, \bibinfo {author}
  {\bibfnamefont {E.~C.}\ \bibnamefont {Garnett}}, \bibinfo {author}
  {\bibfnamefont {M.}~\bibnamefont {Najarian}}, \bibinfo {author}
  {\bibfnamefont {A.}~\bibnamefont {Majumdar}}, \ and\ \bibinfo {author}
  {\bibfnamefont {P.}~\bibnamefont {Yang}},\ }\href {\doibase
  10.1038/nature06381} {\bibfield  {journal} {\bibinfo  {journal} {Nature}\
  }\textbf {\bibinfo {volume} {451}},\ \bibinfo {pages} {163--167} (\bibinfo
  {year} {2008})}\BibitemShut {NoStop}%
\bibitem [{\citenamefont {Li}\ \emph {et~al.}(2014)\citenamefont {Li},
  \citenamefont {Carrete}, \citenamefont {Katcho},\ and\ \citenamefont
  {Mingo}}]{li2014shengbte}%
  \BibitemOpen
  \bibfield  {author} {\bibinfo {author} {\bibfnamefont {W.}~\bibnamefont
  {Li}}, \bibinfo {author} {\bibfnamefont {J.}~\bibnamefont {Carrete}},
  \bibinfo {author} {\bibfnamefont {N.~A.}\ \bibnamefont {Katcho}}, \ and\
  \bibinfo {author} {\bibfnamefont {N.}~\bibnamefont {Mingo}},\ }\href
  {\doibase http://dx.doi.org/10.1016/j.cpc.2014.02.015} {\bibfield  {journal}
  {\bibinfo  {journal} {Comput. Phys. Commun.}\ }\textbf {\bibinfo {volume}
  {185}},\ \bibinfo {pages} {1747 -- 1758} (\bibinfo {year}
  {2014})}\BibitemShut {NoStop}%
\bibitem [{\citenamefont {Zhou}\ and\ \citenamefont
  {Umezawa}(2015)}]{C5CP02255J}%
  \BibitemOpen
  \bibfield  {author} {\bibinfo {author} {\bibfnamefont {W.}~\bibnamefont
  {Zhou}}\ and\ \bibinfo {author} {\bibfnamefont {N.}~\bibnamefont {Umezawa}},\
  }\href {\doibase 10.1039/C5CP02255J} {\bibfield  {journal} {\bibinfo
  {journal} {Phys. Chem. Chem. Phys.}\ }\textbf {\bibinfo {volume} {17}},\
  \bibinfo {pages} {17816--17820} (\bibinfo {year} {2015})}\BibitemShut
  {NoStop}%
\bibitem [{\citenamefont {Mounet}\ \emph {et~al.}(2018)\citenamefont {Mounet},
  \citenamefont {Gibertini}, \citenamefont {Schwaller}, \citenamefont {Campi},
  \citenamefont {Merkys}, \citenamefont {Marrazzo}, \citenamefont {Sohier},
  \citenamefont {Castelli}, \citenamefont {Cepellotti}, \citenamefont {Pizzi},\
  and\ \citenamefont {Marzari}}]{Mounet2018}%
  \BibitemOpen
  \bibfield  {author} {\bibinfo {author} {\bibfnamefont {N.}~\bibnamefont
  {Mounet}}, \bibinfo {author} {\bibfnamefont {M.}~\bibnamefont {Gibertini}},
  \bibinfo {author} {\bibfnamefont {P.}~\bibnamefont {Schwaller}}, \bibinfo
  {author} {\bibfnamefont {D.}~\bibnamefont {Campi}}, \bibinfo {author}
  {\bibfnamefont {A.}~\bibnamefont {Merkys}}, \bibinfo {author} {\bibfnamefont
  {A.}~\bibnamefont {Marrazzo}}, \bibinfo {author} {\bibfnamefont
  {T.}~\bibnamefont {Sohier}}, \bibinfo {author} {\bibfnamefont {I.~E.}\
  \bibnamefont {Castelli}}, \bibinfo {author} {\bibfnamefont {A.}~\bibnamefont
  {Cepellotti}}, \bibinfo {author} {\bibfnamefont {G.}~\bibnamefont {Pizzi}}, \
  and\ \bibinfo {author} {\bibfnamefont {N.}~\bibnamefont {Marzari}},\ }\href
  {\doibase 10.1038/s41565-017-0035-5} {\bibfield  {journal} {\bibinfo
  {journal} {Nat. Nanotechnol.}\ }\textbf {\bibinfo {volume} {13}},\ \bibinfo
  {pages} {246--252} (\bibinfo {year} {2018})}\BibitemShut {NoStop}%
\bibitem [{\citenamefont {Carrete}\ \emph {et~al.}(2016)\citenamefont
  {Carrete}, \citenamefont {Li}, \citenamefont {Lindsay}, \citenamefont
  {Broido}, \citenamefont {Gallego},\ and\ \citenamefont
  {Mingo}}]{carrete2016physically}%
  \BibitemOpen
  \bibfield  {author} {\bibinfo {author} {\bibfnamefont {J.}~\bibnamefont
  {Carrete}}, \bibinfo {author} {\bibfnamefont {W.}~\bibnamefont {Li}},
  \bibinfo {author} {\bibfnamefont {L.}~\bibnamefont {Lindsay}}, \bibinfo
  {author} {\bibfnamefont {D.~A.}\ \bibnamefont {Broido}}, \bibinfo {author}
  {\bibfnamefont {L.~J.}\ \bibnamefont {Gallego}}, \ and\ \bibinfo {author}
  {\bibfnamefont {N.}~\bibnamefont {Mingo}},\ }\href {\doibase
  10.1080/21663831.2016.1174163} {\bibfield  {journal} {\bibinfo  {journal}
  {Mater. Res. Lett.}\ }\textbf {\bibinfo {volume} {4}},\ \bibinfo {pages}
  {204--211} (\bibinfo {year} {2016})}\BibitemShut {NoStop}%
\bibitem [{\citenamefont {Morelli}\ and\ \citenamefont {A.}(2006)}]{Gekappa}%
  \BibitemOpen
  \bibfield  {author} {\bibinfo {author} {\bibfnamefont {D.~T.}\ \bibnamefont
  {Morelli}}\ and\ \bibinfo {author} {\bibfnamefont {S.~G.}\ \bibnamefont
  {A.}},\ }\href {\doibase 10.1007/0-387-25100-6_2} {}edited by\ \bibinfo
  {editor} {\bibfnamefont {S.~L.}\ \bibnamefont {Shind{\'e}}}\ and\ \bibinfo
  {editor} {\bibfnamefont {J.}~\bibnamefont {Goela}}\ (\bibinfo  {publisher}
  {Springer-Verlag New York},\ \bibinfo {year} {2006})\BibitemShut {NoStop}%
\bibitem [{\citenamefont {Qin}\ \emph {et~al.}(2016)\citenamefont {Qin},
  \citenamefont {Qin}, \citenamefont {Fang}, \citenamefont {Zhang},
  \citenamefont {Yue}, \citenamefont {Yan}, \citenamefont {Hu},\ and\
  \citenamefont {Su}}]{C6NR01349J}%
  \BibitemOpen
  \bibfield  {author} {\bibinfo {author} {\bibfnamefont {G.}~\bibnamefont
  {Qin}}, \bibinfo {author} {\bibfnamefont {Z.}~\bibnamefont {Qin}}, \bibinfo
  {author} {\bibfnamefont {W.-Z.}\ \bibnamefont {Fang}}, \bibinfo {author}
  {\bibfnamefont {L.-C.}\ \bibnamefont {Zhang}}, \bibinfo {author}
  {\bibfnamefont {S.-Y.}\ \bibnamefont {Yue}}, \bibinfo {author} {\bibfnamefont
  {Q.-B.}\ \bibnamefont {Yan}}, \bibinfo {author} {\bibfnamefont
  {M.}~\bibnamefont {Hu}}, \ and\ \bibinfo {author} {\bibfnamefont
  {G.}~\bibnamefont {Su}},\ }\href {\doibase 10.1039/C6NR01349J} {\bibfield
  {journal} {\bibinfo  {journal} {Nanoscale}\ }\textbf {\bibinfo {volume}
  {8}},\ \bibinfo {pages} {11306--11319} (\bibinfo {year} {2016})}\BibitemShut
  {NoStop}%
\bibitem [{\citenamefont {Sun}\ \emph {et~al.}(2018)\citenamefont {Sun},
  \citenamefont {Haunschild}, \citenamefont {Polanco}, \citenamefont {Ju},
  \citenamefont {Lindsay}, \citenamefont {KoblmMei~ller},\ and\ \citenamefont
  {Koh}}]{sun2018dislocation}%
  \BibitemOpen
  \bibfield  {author} {\bibinfo {author} {\bibfnamefont {B.}~\bibnamefont
  {Sun}}, \bibinfo {author} {\bibfnamefont {G.}~\bibnamefont {Haunschild}},
  \bibinfo {author} {\bibfnamefont {C.}~\bibnamefont {Polanco}}, \bibinfo
  {author} {\bibfnamefont {J.~Z.-J.}\ \bibnamefont {Ju}}, \bibinfo {author}
  {\bibfnamefont {L.}~\bibnamefont {Lindsay}}, \bibinfo {author} {\bibfnamefont
  {G.}~\bibnamefont {KoblmMei~ller}}, \ and\ \bibinfo {author} {\bibfnamefont
  {Y.~K.}\ \bibnamefont {Koh}},\ }\href {\doibase 10.1038/s41563-018-0250-y}
  {\bibfield  {journal} {\bibinfo  {journal} {Nat. Mater.}\ } (\bibinfo {year}
  {2018}),\ 10.1038/s41563-018-0250-y}\BibitemShut {NoStop}%
\bibitem [{\citenamefont {Fan}\ \emph {et~al.}(2017)\citenamefont {Fan},
  \citenamefont {Sigg}, \citenamefont {Spolenak},\ and\ \citenamefont
  {Ekinci}}]{uniaxial1}%
  \BibitemOpen
  \bibfield  {author} {\bibinfo {author} {\bibfnamefont {D.}~\bibnamefont
  {Fan}}, \bibinfo {author} {\bibfnamefont {H.}~\bibnamefont {Sigg}}, \bibinfo
  {author} {\bibfnamefont {R.}~\bibnamefont {Spolenak}}, \ and\ \bibinfo
  {author} {\bibfnamefont {Y.}~\bibnamefont {Ekinci}},\ }\href {\doibase
  10.1103/PhysRevB.96.115307} {\bibfield  {journal} {\bibinfo  {journal} {Phys.
  Rev. B}\ }\textbf {\bibinfo {volume} {96}},\ \bibinfo {pages} {115307}
  (\bibinfo {year} {2017})}\BibitemShut {NoStop}%
\bibitem [{\citenamefont {Kuwahara}\ \emph {et~al.}(2018)\citenamefont
  {Kuwahara}, \citenamefont {at~Tanusilp}, \citenamefont {Ohishi},
  \citenamefont {Muta}, \citenamefont {Yamanaka},\ and\ \citenamefont
  {Kurosaki}}]{Shimpei}%
  \BibitemOpen
  \bibfield  {author} {\bibinfo {author} {\bibfnamefont {S.}~\bibnamefont
  {Kuwahara}}, \bibinfo {author} {\bibfnamefont {S.}~\bibnamefont
  {at~Tanusilp}}, \bibinfo {author} {\bibfnamefont {Y.}~\bibnamefont {Ohishi}},
  \bibinfo {author} {\bibfnamefont {H.}~\bibnamefont {Muta}}, \bibinfo {author}
  {\bibfnamefont {S.}~\bibnamefont {Yamanaka}}, \ and\ \bibinfo {author}
  {\bibfnamefont {K.}~\bibnamefont {Kurosaki}},\ }\href {\doibase
  10.2320/matertrans.E-M2018804} {\bibfield  {journal} {\bibinfo  {journal}
  {Mater. Trans.}\ }\textbf {\bibinfo {volume} {59}},\ \bibinfo {pages}
  {1022--1029} (\bibinfo {year} {2018})}\BibitemShut {NoStop}%
\bibitem [{\citenamefont {Togo}\ and\ \citenamefont {Tanaka}(2015)}]{phonopy}%
  \BibitemOpen
  \bibfield  {author} {\bibinfo {author} {\bibfnamefont {A.}~\bibnamefont
  {Togo}}\ and\ \bibinfo {author} {\bibfnamefont {I.}~\bibnamefont {Tanaka}},\
  }\href {\doibase https://doi.org/10.1016/j.scriptamat.2015.07.021} {\bibfield
   {journal} {\bibinfo  {journal} {Scripta Mater.}\ }\textbf {\bibinfo {volume}
  {108}},\ \bibinfo {pages} {1 -- 5} (\bibinfo {year} {2015})}\BibitemShut
  {NoStop}%
\bibitem [{\citenamefont {Ziman}(1960)}]{ziman}%
  \BibitemOpen
  \bibfield  {author} {\bibinfo {author} {\bibfnamefont {J.~M.}\ \bibnamefont
  {Ziman}},\ }\href {https://cds.cern.ch/record/100360} {}International series
  of monographs on physics\ (\bibinfo  {publisher} {Clarendon Press},\ \bibinfo
  {address} {Oxford},\ \bibinfo {year} {1960})\BibitemShut {NoStop}%
\bibitem [{\citenamefont {Qiao}\ \emph {et~al.}(2014)\citenamefont {Qiao},
  \citenamefont {Kong}, \citenamefont {Hu}, \citenamefont {Yang},\ and\
  \citenamefont {Ji}}]{qiao2014high}%
  \BibitemOpen
  \bibfield  {author} {\bibinfo {author} {\bibfnamefont {J.}~\bibnamefont
  {Qiao}}, \bibinfo {author} {\bibfnamefont {X.}~\bibnamefont {Kong}}, \bibinfo
  {author} {\bibfnamefont {Z.-X.}\ \bibnamefont {Hu}}, \bibinfo {author}
  {\bibfnamefont {F.}~\bibnamefont {Yang}}, \ and\ \bibinfo {author}
  {\bibfnamefont {W.}~\bibnamefont {Ji}},\ }\href
  {http://dx.doi.org/10.1038/ncomms5475} {\bibfield  {journal} {\bibinfo
  {journal} {Nat. Commun.}\ }\textbf {\bibinfo {volume} {5}},\ \bibinfo {pages}
  {4475} (\bibinfo {year} {2014})}\BibitemShut {NoStop}%
\bibitem [{\citenamefont {Watson}\ and\ \citenamefont
  {Parker}(1999)}]{bulkPbO}%
  \BibitemOpen
  \bibfield  {author} {\bibinfo {author} {\bibfnamefont {G.~W.}\ \bibnamefont
  {Watson}}\ and\ \bibinfo {author} {\bibfnamefont {S.~C.}\ \bibnamefont
  {Parker}},\ }\href {\doibase 10.1021/jp9841337} {\bibfield  {journal}
  {\bibinfo  {journal} {J. Phys. Chem. B}\ }\textbf {\bibinfo {volume} {103}},\
  \bibinfo {pages} {1258--1262} (\bibinfo {year} {1999})}\BibitemShut {NoStop}%
\bibitem [{\citenamefont {Watson}(2001)}]{bulkSnO}%
  \BibitemOpen
  \bibfield  {author} {\bibinfo {author} {\bibfnamefont {G.~W.}\ \bibnamefont
  {Watson}},\ }\href {\doibase 10.1063/1.1331102} {\bibfield  {journal}
  {\bibinfo  {journal} {J. Chem. Phys.}\ }\textbf {\bibinfo {volume} {114}},\
  \bibinfo {pages} {758--763} (\bibinfo {year} {2001})}\BibitemShut {NoStop}%
\bibitem [{\citenamefont {Skoug}\ and\ \citenamefont
  {Morelli}(2011)}]{lonepair}%
  \BibitemOpen
  \bibfield  {author} {\bibinfo {author} {\bibfnamefont {E.~J.}\ \bibnamefont
  {Skoug}}\ and\ \bibinfo {author} {\bibfnamefont {D.~T.}\ \bibnamefont
  {Morelli}},\ }\href {\doibase 10.1103/PhysRevLett.107.235901} {\bibfield
  {journal} {\bibinfo  {journal} {Phys. Rev. Lett.}\ }\textbf {\bibinfo
  {volume} {107}},\ \bibinfo {pages} {235901} (\bibinfo {year}
  {2011})}\BibitemShut {NoStop}%
\bibitem [{\citenamefont {Qin}\ \emph {et~al.}(2018)\citenamefont {Qin},
  \citenamefont {Qin}, \citenamefont {Wang},\ and\ \citenamefont
  {Hu}}]{qin2018}%
  \BibitemOpen
  \bibfield  {author} {\bibinfo {author} {\bibfnamefont {G.}~\bibnamefont
  {Qin}}, \bibinfo {author} {\bibfnamefont {Z.}~\bibnamefont {Qin}}, \bibinfo
  {author} {\bibfnamefont {H.}~\bibnamefont {Wang}}, \ and\ \bibinfo {author}
  {\bibfnamefont {M.}~\bibnamefont {Hu}},\ }\href {\doibase
  10.1016/j.nanoen.2018.05.040} {\bibfield  {journal} {\bibinfo  {journal}
  {Nano Energy}\ }\textbf {\bibinfo {volume} {50}},\ \bibinfo {pages} {425 --
  430} (\bibinfo {year} {2018})}\BibitemShut {NoStop}%
\end{thebibliography}%
\end{document}